# Exploratory Analysis of a Terabyte Scale Web Corpus

Vasilis Kolias[1], Ioannis Anagnostopoulos[2], Eleftherios Kayafas[1]

[1] *National Technical University of Athens, 9 Heroon Polytechneiou st. 15780 Zografou campus, {vkolias@medialab.ntua.gr, kayafas@cs.ntua.gr}*
[2] *University of Thessaly, 2-4 Papassiopoulou st. 35100 Lamia, janag@ucg.gr*

*Abstract* – In this extended abstract we present a preliminary (and at this point incomplete) analysis over the largest publicly accessible web dataset: the Common Crawl Corpus. We measure nine web characteristics from two levels of granularity using MapReduce and we comment on the initial observations which will be concluded with the final version of the paper. To the best of our knowledge two of the characteristics, the language distribution and the HTML version of pages have not been analyzed in previous work, while the specific dataset has been only analyzed on page level.

## I. INTRODUCTION

The rapid and continuous growth of the Web over the recent past makes it the largest publicly accessible data source in the world. The Web has many unique characteristics, which make finding useful information and discovering knowledge a challenging yet fascinating task. Although its exact size cannot be accurately measured, the Web can be studied through the other properties it exposes, like content dynamism and interconnectedness.

In order to study the Web as a whole, a significant amount of its content has to be collected. The automated process of traversing the web by repeatedly following hyperlinks in order to retrieve content for further processing, is known as web crawling [1]. While the prominent use of crawling is to build search engine indexes, harvested content can also be used for archiving [2], data mining and analytics [3], linguistic analysis, or for triggering services [4].

During the last decade, enormous corpora of web content have been created by companies and corporations which constitute the base of their services. Access to these datasets is restricted and any kind of study is limited within their premises. At the same time, open collections for information retrieval research have been made publicly available - mainly by the academia - for a variety of domains ranging from blogs to newswire text. And although they adequately served many research purposes, for other research questions they are sometimes too small, other times unbalanced, or they are focused on specific criteria for the task at hand.

In this paper we conduct an exploratory analysis of the largest publicly available dataset for the Web: the Common Crawl corpus. The crawl is gathered and maintained by the Common Crawl Foundation [5] with the aim to provide access to Web information for everyone. The crawl is hosted on Amazon Simple Storage Service (S3) [6] in a hierarchically organized manner, with directory segments containing files in the ARC and WARC formats [7][8], sorted by date. In this paper we analyze the 2012 corpus, though, as of this writing, the Common Crawl offers datasets for the years 2010 through 2014. The 2012 corpus contains almost 4 billion web pages containing 130 billion links.

Perhaps the most important aspect of a crawling process is the strategy employed by the crawlers. The crawler used for the Common Crawl corpus is based on a breadth-first strategy along with heuristics to detect and avoid spam pages, duplicates or empty pages. In a breadth-first strategy the crawler starts with an initial set of URLs known as the seed, visits them, extracts the URLs in their content, adds them to a queue which is known as the frontier and repeats this process until the queue is empty, or other criteria have been met. This strategy in general, leads to an exhaustive collection of locally connected sub components of the web graph, while possibly leaving other interesting material in other parts of the graph unvisited. Together with the heuristics and the initial seed of the Common Crawl crawler, this strategy has potentially affected the paths visited and the distribution of host sizes, since biases towards popular websites may have been introduced. While the Common Crawl corpus gets bigger each year, to claim that it is a representative sample of the Web is fairly subtle. The dataset is essentialy nothing more than a large "snapshot" of the Web from the period of the crawling process. Nevertheless, the size of the dataset alone constitutes an appealing factor for research initiatives that may lead to useful conclusions about the Web in its entirety.

## II. RELATED WORK

Since its establishment, the Web has been studied from various perspectives and with different objectives. The first work on the structure of the

Web was published back in 2000 [9]. Their work was based on two crawls of 200 million pages and 1.5 million links, with the second crawl serving for validating the first. In their findings, they introduced a structural model of the Web known as the bowtie model, which is a division of the Web into various components, based on how they are connected. Along with page in-degree and out-degree, they presented the distribution of the sizes of strongly connected components in their model, all of which follow power laws. Various other studies focused on the structure of the so called "national" Web domains, which consist of all Web sites that are registered at a domain inside the assigned country code or that are hosted at an IP that belongs to a segment assigned to the country been presented. Works [10-15] present findings on crawls made by different crawlers and on different parts of the Web, which provide different snapshots of the web graph. A more detailed paper about the size of the components of the bowtie model can be found in [16]. They analyzed four crawls gathered between 2001 and 2004 by different crawlers with different parameters and they concluded that several properties of web crawls are dependent on the crawling process. Finally, an in depth comparison of the latest findings on the web structure with previous works, is done in [17]. They confirm the existence of a giant strongly connected component, but they strongly emphasize that it is strongly dependent on the crawling process. Their most important finding however is that the distributions of indegree, outdegree and sizes of strongly connected components are not power laws something that contradicts to the evidence found throughout the literature up to now. This work also used the Common Crawl dataset.

Along with its structure, other characteristics of the Web are presented in [18]. This work is essentially a side-by-side comparison of the results of 12 Web characterization studies, comprising over 120 million pages from 24 countries. Their results include various levels of detail at which different aspects are presented, while they separate their findings between contents, links, and technologies. Finally a technical report also presented the main characteristics of the Common Crawl 2012 dataset [19].

### III. SETUP AND METHODOLOGY

As already mentioned, the 2012 Common Crawl corpus contains almost 4 billion web pages which occupy 210 terabytes of data. The only feasible approach to harness such a large dataset is to use divide and conquer models such as MapReduce. MapReduce [20] is a programming model and an associated implementation for processing and generating large datasets. It consists of two stages. In the first stage, a computation specified by the programmer called map, is applied all over the dataset which is initially split into input records. These computations occur in parallel and they emit intermediate results which are aggregated by the second stage: another programmer specified operation known as reduce. The execution framework coordinates the actual processing, thus leaving to the programmer, only the implementation of the map and reduce functions. In this paper we used the software implementation of MapReduce model and framework called Hadoop running on Elastic Map Reduce (EMR) Amazon web service [21].

In order to handle the dataset efficiently we divided the 2012 Common Crawl dataset into smaller manageable subsets and processed them individually in two steps. In the first step we analyzed the web pages from the subsets and reduced them into a single intermediate result set. In the second step we aggregated the intermediate results into the final publishable results. Both of these steps were plain mapreduce programs implemented in Java and their execution was done in a 50 node EMR cluster of m3.xlarge instances running for 1600 instance hours.

As stated in [22] the Web can be studied at several levels of granularity. Beginning from a single byte, one can analyze the Web on the level of words, pages, sites, domains, top level domains, national web domains and finally as a whole. The level of study depends on the sample of the Web at hand and throughout the literature three types of sampling have been observed: a) complete crawls of a single website, b) random samples of the whole Web and c) large samples of specific communities. While the total number of exhaustively crawled websites present in the Common Crawl corpus is unknown, the diversity of languages and domains inside the corpus is evident. Therefore, one can argue that to some extent the Common Crawl corpus can be seen as a superset of all the aforementioned levels of granularity and types of sampling.

### IV. ANALYSIS OF WEB CHARACTERISTICS

In this paper we analyzed a a rather small portion of the Common Crawl corpus consisting of 176773760 resources. Almost 90% of this subset consists of html pages, while the rest is comprised mostly of pdf, xml, css, jpeg and javascript files. Table 1 presents the top-10 resource types sorted by their frequency.

The rest of the analysis is focused on two granularity levels: the page level and the website level.

*A. Page Level Analysis*

We begin our investigation by examining four characteristics on the page level, namely the page size, the page language, the page age and the HTML version used for the page. We have accounted the page size as the size of the full HTTP response of the request for that page since if we relied on the Content-Length field value of the response headers we would not have values for the 45% of the responses. The rest of the values are summarized in Figure 1.

The next characteristic is the page language. For language detection we relied on a Naive Bayes classifier [23] which updates the posterior probabilities of the supported languages by a set of predefined n-gram probabilities for each language until they reach a specified threshold.

Table 1. The MIME type distribution of resources.

| MIME Type | Abs. Freq. | Rel. Freq |
|---|---|---|
| text/html | 160.095.983 | 90.5% |
| application/pdf | 5.705.217 | 3.2% |
| text/xml | 5.581.802 | 3.1% |
| text/css | 1.158.062 | 0.6% |
| image/jpeg | 684.593 | 0.3% |
| application/x-javascript | 680.597 | 0.3% |
| application/msword | 435.147 | 0.2% |
| text/plain | 400.444 | 0.2% |
| application/javascript | 341.496 | 0.1% |
| application/rss+xml | 211.517 | 0.1% |

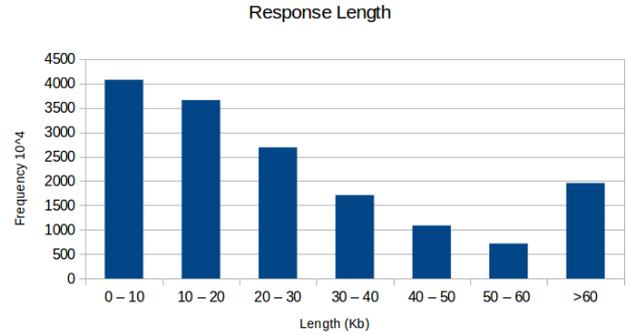

Figure 1. Response Length Distribution

Figure 2 presents the top 10 languages for the subset of the corpus we investigated. We can clearly see that English is the dominant language but we also see the absence of Asian languages, something that must be further investigated for the remaining dataset. Also the classifier could not detect the language for a 5.7% of the pages. Another point to notice is that relying on the Content-Language field of the HTTP response is not safe since 98% of the values where missing. Therefore in order to accurately determine the language of a page, language detectors have to be employed.

The next characteristic we examine is the page age. Age is not easily estimated since values Date and Last-Modified fields of the HTTP response headers are not always given. In fact 64% of the responses did not include one or both of the two fields. Also a very insignificant amount of pages provided negative age. The rest of the values can be seen in Figure 3. Age is another characteristic that needs to be estimated with another method, since the HTTP response headers are once more proven unreliable.

Another characteristic we examine is the HTML version that was used for each page. Figure 4 presents the distribution of the various HTML versions found in the subset. We see that XHTML is the dominant version in the subset but we also have to note that a) 20% of the pages did not provide any version and b) the pages in HTML5 did not necessarily include any HTML5 specific code. However we can safely claim that in this subset, HTML5 has not been widely adopted.

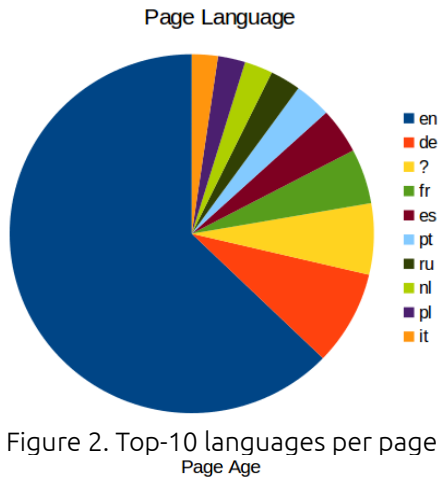

Figure 2. Top-10 languages per page

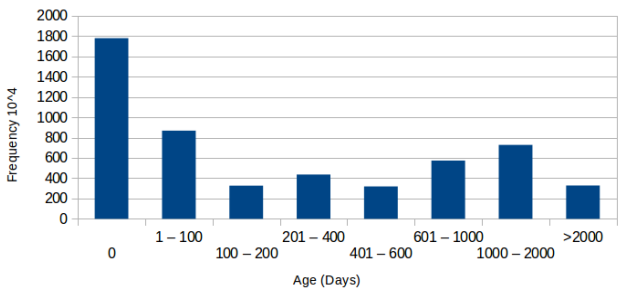

Figure 3. Age Distribution

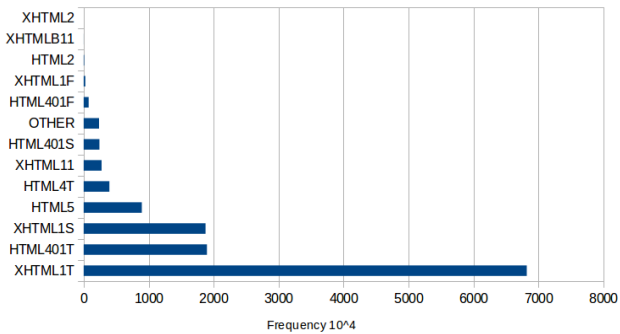

Figure 4. HTML Version Distribution

We further investigate two structural properties of pages: the in-degree and the out-degree. One of the most important features of the Web is its connectedness, i.e. the ability of one web page to link with another web page via special markup called hyperlinks. The cardinality of the set of pages linking to a particular page is called the in-degree of the particular page and the cardinality of the set of pages linked by the specific page is called the out-degree of the page. Figure 5 and 6 depict the out-degree and in-degree respectively. Throughout the literature the two measures have been reported to follow power law distributions. We observe that the total references and in-links for the portion of the dataset we examine, lie between 0 and 500 and that the distributions on both in-degree and out-degree drops quickly. Both of these observations are influenced by the small size of the subset examined and further investigation is required on the whole dataset.

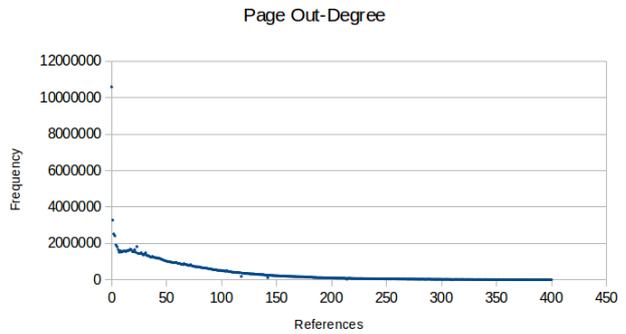

Figure 5. Page Out-degree

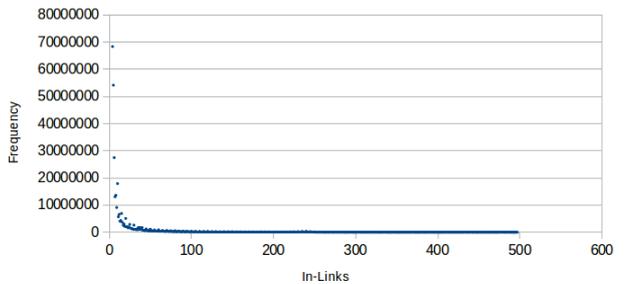

Figure 6. Page In-Degree

### B. Site Level Analysis

We begin our investigation on website level by exploring the page distribution and language per website. In Figure 7 web observe the distribution of pages per website. 68% of the websites seem to be comprised of a single page. This amount does not seem realistic and it can be attributed to the small size of the subset. It is highly possible that for most of these websites the rest of their pages lie in the rest of the dataset.

The top-10 distribution of languages is quite the same with that of web pages (Figure 8). The dominant language here is also English, but we notice some modifications on the proportions of the rest. This is due to the fact that we essentially ignore the number of pages per website. As far as the language, another interesting observation is the number of different languages a website supports. 94% of the websites are written in a single language while the rest are multilingual.

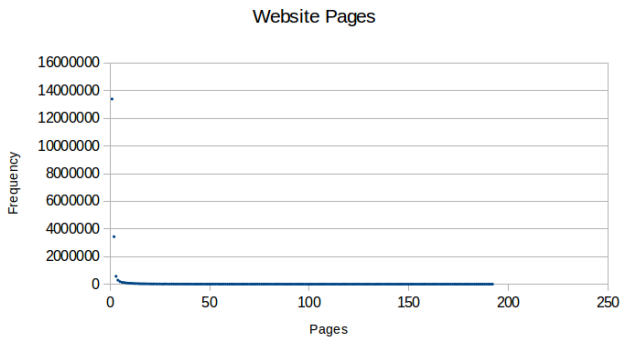

Figure 7. Pages per website

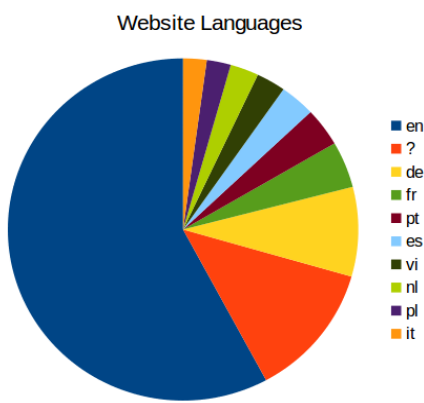

Figure 8. Top-10 Language distribution per website

Finally, the out-degree of a website (Figure 9), which was calculated by counting the domain of the URLs found on the references of all the pages of the website, quite resembles the distribution of the out-degree of single pages. Again safer conclusions about the distribution can be drawn by taking into account the dataset in its entirety.

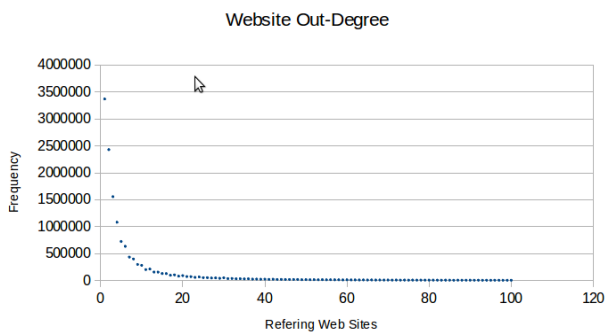

Figure 9. Website out-degree

## V. CONCLUSIONS

In this extended abstract we presented an initial exploratory analysis on the largest publicly available web corpus, the Common Crawl. Although we examined only a fraction of the dataset, some initial useful measurements were shown. A thorough and complete investigation against the whole Common Crawl corpus can lead to useful conclusions and most of all, it can prove the suitability of the dataset for further research.